\documentclass{ifacconf}

\usepackage{graphicx}      % include this line if your document contains figures
\usepackage{natbib}        % required for bibliography
\usepackage{url}
\usepackage{color}
\usepackage{amsmath}

\begin{document}
\begin{frontmatter}

\title{On the Dynamics of Comets in \\
Extrasolar Planetary Systems
%  Paper Title\thanksref{footnoteinfo}
} 

\author[First]{R. Dvorak} 
\author[First]{B. Loibnegger} 
\author[Third]{M. Cuntz}

\address[First]{Institute of Astronomy, 
   University of Vienna, A-1180 Vienna, Austria } %(e-mail: rudolf.dvorak@univie.ac.at).
\address[Third]{Department of Physics, 
   University of Texas at Arlington, Arlington, TX 76019, USA} %(e-mail: cuntz@uta.edu)

\begin{abstract}                % Abstract of not more than 250 words.

Since very recently, we acquired knowledge on the existence of comets in extrasolar
planetary systems.  The formation of comets together with planets around host
stars now seems evident.  As stars are often born in clusters of interstellar clouds,
the interaction between the systems will lead to the exchange of material at the edge of the clouds.
Therefore, almost every planetary system should have leftover remnants as a result of planetary
formation in form of comets at the edges of those systems. These Oort clouds around stars are
often disturbed by different processes (e.g., galactic tides, passing stars, etc.), which
consequently scatter bodies from the distant clouds into the system close to the host star.
Regarding the Solar System, we observe this outcome in the form of cometary families.
This knowledge supports the assumption of the existence of comets around other stars.
In the present work, we study the orbital dynamics of hypothetical exocomets, based on
detailed computer simulations,  in three star-planet systems, which are: HD~10180, 47~UMa,
and HD~141399.  These systems host one or more Jupiter-like planets, which change
the orbits of the incoming comets in characteristic ways.

\end{abstract}


\end{frontmatter}
%===============================================================================

\section{Introduction}

For thousands of years the non-predictability of the appearance of comets
contrary to the motion of the planets, the Sun, and the Moon led astronomers to
the assumption that they originate within the Earth's atmosphere. Thanks to
Tycho Brahe, who measured their parallax at the end of the 16th century, it could be established
that these objects originate as far away as at least 4 times the distance to our Moon. Using
Newton's laws, Sir Edmond Halley was even able to determine the orbit of a
comet and predicted its reappearance after a period of 76 years.
Before that comets were thought to be heavenly 
omens for disaster and catastrophes on Earth.
Only recently, based on spectroscopic measurements, it was
found that their role in the development of life on Earth is essential:
although the Earth may have been formed with some water in the crust and the
mantle already, a considerable part of it (as well as molecules of biochemical significance)
has been transported onto the surface of the Earth afterward through asteroids and mainly
comets.  Comparing the estimated water content of comets within the Oort
cloud\footnote{A reservoir of billions of comets in the Kuiper belt and in the Oort cloud thousands of au away from the Sun.}
the ratio between the water in this cloud compared to the Earth's water in the oceans is estimated to be on the
order of $10^4$! Still, there is an ongoing discussion whether the water in the
oceans and the one observed in comets is different concerning the ratio of
hydrogen to deuterium {\citep[e.g.,][]{Raymond2017}}\footnote{So far, the deuterium/hydrogen (D/H) ratio has been
measured in 11 different comets, covering both types, but only one shows similarity with Earth: the Jupiter-family Comet 103P/Hartley 2.
However, as reported in {\it Science}, {\it Rosetta} found that the D/H ratio is more than three times higher
than the values found in Earth’s oceans and on Hartley 2.  Hence, this question is
far from being settled!}.  Considering the large number of comets which entered the inner
Solar System and collided with the Earth during the last billions of years, it is worth
to pursue statistical analyses for these events in extrasolar planetary systems as well.

Our paper is structured as follows:
In the next chapter, we explain how comets could have formed around other stars.
Thereafter, we discuss the observational evidence for comets in extrasolar systems
together with a list of exoplanetary systems with gas giants akin to Jupiter and
Saturn.  In the main chapter, we show the results of simulations for the three systems
HD~10180, 47~UMa, and HD~141399.
Finally, we discuss what kind of consequence our results would have on possible
water transport to possible terrestrial planets in stellar habitable zones.

\section{Extrasolar Oort Clouds}\label{sect:extrasolarortcl}

\cite{Stern1987} explored how many comets from a so-called extrasolar Oort cloud
may have hit the Earth as well as other terrestrial planets or Jupiter.
According to his results Earth may have suffered from a cometary impact
originating from an Oort cloud around a star other than the Sun every $10^7$
years\footnote{In this article the author distinguished between comet nuclei and coma.}.
In comparison, in this article the impact rate for Solar indigenous comets was found
to be three orders of magnitude higher, which was estimated according to previously computed
probability rates \citep{Whipple1950}. 

We mention this because in this article the existence of extrasolar Oort clouds is assumed,
a concept previously considered as well by other
authors \cite[e.g.,][]{Kaib2008,Kaib2011,Brasser2006,Brasser2007}.
In fact, the formation of the Sun was not an isolated event occurring in an
interstellar cloud but took place in an extended cloud where many other stars were born as well.
This process was accompanied by the formation of objects such as comets containing 
a large array of molecules, including many molecules of astrobiological relevance.
Looking at the
ingredients of known comets, as, e.g.,
67P/Churyumov-Gerasimenko (visited by the {\it Rosetta} mission), the dust consists of almost
50$\%$ organic molecules and a notable part of water\footnote{The expression of `dirty snowball' was introduced by \cite{Whipple1950}.}.
Thus, huge amounts of small icy bodies were created in the early disk around a star.
Due to perturbations from nearby stars and other interstellar processes, many (or most)
outer objects of this extrasolar Oort cloud were ejected and lost in the interstellar space as free interstellar comets \citep{Stern1987}.
Alternatively, they were captured by the Sun or other stars, or were able to travel to the inner part of the Solar System as
previously discussed.

Thus, comets should be considered a by-product of planet formation.  This leads to the assumption that in
most, if not all planetary system there should be cometary activity, despite the fact that the term ``exocomets"
has been coined only very recently.

The dynamics of scattering of planetesimals by a planet has been investigated, e.g., by \cite{Wyatt2017},
who explored the scattering outcome depending on the mass and semi-major axis of the planet.  Object trajectories
resulted in accretion, ejection, remaining, escape, or placement in an Oort cloud or a depleted Oort cloud analogue. 

Here we will present the dynamical studies pursued by our group through investigating the outcome of
planet--planetesimal scattering in single star systems.  Therefore, the interaction between test particles
distributed in a disk about the host star and (at least) one massive planet will be explored.  Statistics
for sets of models will be shown.  This will include studies of outwardly scattered objects from
Solar System analogues to the Kuiper belt or the Oort Cloud.  Semi-major axis, eccentricity, perihel, aphel,
inclination, and orbital period after the integration of each test particle are determined and
statistical analyses are performed.

\cite{Rappaport2018} presented observational evidence for exocomet transits in
the data of the \textit{Kepler} star KIC~3542116. The signals show a distinct
asymmetric shape and duration of 1~day or shorter, and have been fitted with a
simple dust-tail model.  The model yields that these comets have approximately the
mass of Halley's comet.  The inferred speeds during the transits are in the range of
35 to 50 km~$\mathrm{s^{-1}}$ for the deeper transits, and 75 to 90
km~$\mathrm{s^{-1}}$ for the more shallow and narrow transits. Similar
observations of short-time events in high-resolution spectra of the Ca~II~K
line of $\phi$~Leo have been presented by \cite{Eiroa2016} who compare those results
to the ones found for $\beta$~Pic \citep{Beust1990,Kiefer2014}.  Based on
observational data from the Atacama Large Millimeter/submillimeter Array (ALMA)
\cite{Matra2017} presented evidence for exocometary gas released within a Kuiper Belt analogue
at $\sim$136~au around the 440~Myr-old Formalhaut system. The amount of CO observed implies
that it originates from exocometary ices within the belt and is consistent with observations
of Solar System comets.  

\section{Evidence of the Existence of Extrasolar Comets}

\cite{Welsh2015} presented observations of absorption line profiles for
Ca~II~K (3933 $\mbox{\AA}$) for 15 A-type and two B-type stars with known
debris disks.  They found short-term absorption variation in the line profiles
in one late B-type and four A-type stars.  The observed
spectra contain a rotationally-broadened absorption line at 3933~$\mbox{\AA}$,
which is typical for circumstellar gaseous disks co-rotating near to the
stellar radial velocity of their host star.  Additionally, they identified transient
weak absorption lines occasionally appearing to be red-shifted or blue-shifted by
tens of km~s$^{-1}$.  These features are believed to be caused by falling
evaporating bodies (FEBs) --- analogues to comets in the Solar System --- which
evaporate material when approaching their host stars.  As the circumstellar
K-line may often vary in absorption strength itself, they applied the method
of comparing the night-to-night equivalent width (EW) of each observed star
in order to evaluate if the observed change is accompanied by a significant change
in the main circumstellar line.  They concluded that the observational results
are consistent with the presence of FEBs.

Moreover, \cite{Welsh2015} also discussed the systems of HD~64145, HD~56537,
HD~58647 (late B-type), HD~108767, HD~9672, HD~80007, HD~109573, and
$\beta$~Pictoris in detail.  The key findings of this paper read as follows:

The stars hosting FEBs are on average 70 Myr younger than the non-FEB hosting stars
--- although one has to consider that obtaining accurate ages for A-stars is still
problematic.  They also could not find a statistically significant difference between
stars based on their metallicity or chemical peculiarity.  In the observed sample
the FEB hosting stars are of an earlier type which, as argued by the authors, could
be due to stellar activity.  The detection of an FEB event itself strongly depends
on the viewing angle of the disk (which must be close to edge-on) and the observational
time frame (i.e., lower levels of activity need longer timespans of observation).
They conclude that the detection of FEB activity can also be seen as evidence
for the presence of associated exoplanets.
FEB absorption in the Fe~I line is only detected if the similar event is observed
with an equivalent width value of $>50$ m$\mbox{\AA}$ in the Ca II K-line profile.
This relationship between the Fe~I and Ca~II circumstellar absorption in
young disk systems is still not fully understood as discussed in more detail by
\cite{Welsh2016}.

Rotational velocity, mid-IR excess, or chemical peculiarity do not differentiate
between stars with or without FEB activity.  Stellar age does not seem to be
a significant factor either. 
While the above described paper investigated a sample of stars in order to arrive at
conclusions about parameter revealing hints on the presence of exocomets, \cite{Marino2017}
focused on ALMA observations of a single star system, i.e., $\eta$~Corvi.  It is 1.5 Gyr old
system with a two-component debris disk.

The underlying problem was finding a theoretical explanation for the hot dust disk,
which cannot be explained by a collisional cascade \textit{in situ}.
However, they were able to present the analysis for the first ALMA band observations
(7 runs at 340 GHz); those corresponded to observations of the continuum dust emission of
$\eta$~Corvi's outer belt at a wavelength of 0.88 mm.  At this wavelength the continuum is
dominated by $\sim$0.1 -- 10.0~mm sized dust grains for which radiation forces are negligible.
The outer disk is detected with a peak radius of 150~au and a radial width of over 70~au. 
In order to receive estimates for the different disk parameters, they modeled the observed values
to four different disk models. The first model consisted of a simple belt with radial and vertical
Gaussian mass distribution peaked at 150~au with a FWHM (full width half maximum) of 44~au fit best.
The comet-like composition and short life time of the observed belt led to the conclusion
that it is fed from the outer belt via scattering by a chain of planets. 

They propose the following scenario as being responsible for the observations:

Volatile-rich solid material from the outer belt is scattered inwardly via a chain of planets.
This icy material starts to sublimate and loses part of its volatiles at specific ice lines, thus
producing the CO feature observed at $\sim$20 au.  The authors deduced a probable mass distribution
for the chain of planets responsible for the scattering, which should be close to flat between
3 and 30 Earth-masses.  The inwardly scattered material could also explains an \textit{in situ}
collisional cascade or a collision with a planet of 4-10 Earth-masses located at $\sim$3~au
(sweet spot of the system) releasing large amounts of debris and causing the asymmetric structure
revealed in the observations. 

In our investigation we concentrate on the orbital dynamics of exocomets as
a theoretical study.  We show results of computer simulations for three
systems, which are: HD~10180, the 47~UMa system, and HD~141399.  These systems
are known to host one or more Jupiter-mass planets, which are expected to
change the orbits of the incoming comets in characteristic ways.

\section{Three examples about the dynamics of extrasolar comets}

\subsection{Numerical Setup}

Guided by the structure of the Solar System and our knowledge about planet formation, we assume that most, if not all stellar systems harboring planets might also harbor
an Oort-type cloud of comets away far from the star, resulting from the scattering processes occurring during the process of planet formation.
These comets are expected to be gravitationally disturbed by, e.g., a passing star or by galactic tidal forces leading to highly eccentric cometary trajectories and allowing the small objects to enter the planetary region close to the star.  The underlying injection mechanisms have been studied, e.g., by \cite{Fouchard2007b}, \cite{Fouchard2017}, \cite{Fouchard2018} and \cite{Rickman2008}. 

We have not implemented the forces due to galactic tides and/or passing stars in our models as the 
initial semi-major axes of the comets are rather small and the eccentricities are rather large.
Thus, the planets in the system will play the main role in the evolution of the cometary trajectories. 
Furthermore the timespan of our integrations is only 1~Myr, which is too short for the galactic tides 
to significantly affect the dynamics.

We consider tens of thousands of fictitious comets assumed as massless, which we distributed evenly in a sphere with initial conditions as follows:

\begin{itemize}
\item[-] $80~{\rm au} < a < 200~{\rm au}$  with $\delta$a = 10 au
\item[-] $0.915 < e < 0.99$ with $\delta$e = 0.005
\item[-] $0 < i < 180$ with $\delta$i = 10
\end{itemize}

As the comets have been assumed to originate from an Oort-cloud analogue, they have been placed in highly eccentric orbits.
The total integration time was set to 1 Myr.  Whenever a comet was ejected from the system or underwent a collision,
another comet was inserted with the same initial conditions as the previous comet.  Since at that time of insertion
the configuration of the planets is different from the previous setting, the newly inserted comet will undergo different dynamics
within the system.

We track the comet's osculating elements (e.g., semi-major axes and eccentricities) and determine if it is captured into a
long-period or short-periodic orbit.  This happens mostly for comets entering the system in the same plane as of the planets
(see Figure~\ref{fig:1.1}). For some systems (as done for 47~UMa), we monitor the collisions in order to estimate the
possible amount of water transported to an Earth-mass planet in the system's habitable zone.  The ejections and close encounters
of the comets with the planets in the systems are examined as well. 

The integration of the equations of motion was done with the Lie-integration method with adaptive step-size control
\citep{Hanslmeier1984, Eggl2010}. 

The initial conditions for the comets are the same for all three planetary systems considered. As the comets
with smallest semi-major axes are started at 80 au, they are far enough outside the orbit of the outermost planet
of each system. Nevertheless, the noticeably different orbits of the planets in the systems in combination
with the initial conditions for the comets are expected to impact the results.  A relatively large semi-major axis of
the outermost planet is expected to lead to a higher possibility of interaction with the comets and may thus affect the results 
in a way that more collisions and/or captures might occur.  As mentioned in section~\ref{sect:extrasolarortcl}, one can assume 
that the cometary reservoirs have been formed inside a gas cloud with no preferences of inclination relative to the planetary orbits 
of the systems; consequently, the initial conditions were chosen as described.

Due to the lack of knowledge on the inclination of the planets in our study, 
all inclinations were set to randomly small values ($<$ 1$^{\circ}$).

\subsection{HD~10180}

\subsubsection{The System}
\vspace*{0.2cm}
\noindent HD~10180 is a Sun-like star. Its parameters are: $T_\mathrm{eff}$=5910~K, $M$ = 1.06 $M_{\odot}$,
combined with an age of about 4.3 Gyr \citep{Lovis2011}.  The system consists of at least five Neptune-mass planets
and one inner Earth-mass planet (HD~10180~b). \cite{Tuomi2012} found two statistically significant signals corresponding
to two additional super-Earth-mass planets located close-in. Thus, HD~10180 is a tightly packed system with 6,
or possibly 9 planets.  The most massive planet, HD~10180~h, is located beyond the other planets at a distance of
approximately 3.4~au.   For our numerical investigation we only took the four outer planets into account.
The properties of the planets are shown in Table~\ref{t0}.

\begin{table}
	\centering
\begin{tabular}{c|rlc}
\hline
Name & $a$ (au) & $e$ &  $m$ ($\mathrm{m_{Jupiter}}$) \\
\hline
HD~10180~e &  0.270   & 0.0260  &   0.0789 \\
HD~10180~f &  0.49220 & 0.1350  &   0.0752	\\
HD~10180~g &  1.4220  & 0.00010 &   0.0673 \\
HD~10180~h &  3.40    & 0.080   &   0.2026 \\
\hline
\end{tabular}
\caption{Properties of the planets in the system of HD~10180. For details and updates see \protect\url{http://exoplanet.eu/}.}\label{t0}
\end{table}

\subsubsection{Results}
\vspace*{0.2cm}
\noindent 
Similar to the role of Jupiter in the Solar System, which is mostly responsible for the scattering
of small objects (especially comets) originating from the outer parts of the system, in HD~10180 the outermost
and most massive planet HD~10180~h is able to change the orbits of incoming comets most profoundly.
{A possible outcome of gravitational interaction of a comet with the planet is a capture of that small body
in an orbit with a relatively low semi-major axis and small eccentricity.
Such comets may stay in the system for longer time spans akin to the Jupiter-family or Halley-type comets
in the Solar System. Figure~\ref{fig:1.1} conveys the probability for an incoming comet to be captured in
such an orbit depending on the initial conditions. The three peaks visible at about 30$^{\circ}$, 90$^{\circ}$, and 150$^{\circ}$
(green lines in Figure~\ref{fig:1.1}) deserve more detailed investigations; they are most probably
due to the Kozai mechanism.

Nevertheless, the most
probable resulting trajectory after a close encounter is one leading to the escape of the comet from the system.
Figure~\ref{fig:1.2} shows the number of new comets injected into the system; this number agrees with the number
of comets ejected from the system.  It is found that the number of ejected comets increases for comets entering
the system with relatively large initial eccentricities and relatively small initial inclinations.  These initial
parameters ensure that the comet is able to enter the system's inner region and experiences relatively
early close encounters resulting in its escape.

The interaction of a comet with the most massive planet HD~10180~h, leading to either a stable close-in orbit,
a quasi-stable orbit, or a cometary escape, is revealed in an example given in Figure~3 of \cite{Loibnegger2017}.
Here over a period of approximately 1.5 Myr, a comet experiences a stable period.  The orbit is captured at
a perihelion distance of about 5~au before it undergoes a close encounter with the most massive planet
leading to pronounced changes in its orbit.

Figure~\ref{fig:1.3} and \ref{fig:1.4} show the outcome of the scattering process when leading to cometary
captures.  Undergoing close encounters with a planet can lead to an exchange of sufficient amounts of
angular momentum that puts the comet in an orbit with a reduced semi-major axis and eccentricity.
The figures show the cometary orbital elements after an integration time of 1~Myr. We could
not find a clear distinction between long-period and short-periodic comets, as given in our Solar System.
However, it is evident that there are some comets captured in orbits of relatively low eccentricities ($e < 0.3$)
and semi-major axes ($a < 10$~au).  The color coding shows the inclination of the orbit of the captured comets.
Nevertheless, one can observe that most of the comets are captured in highly eccentric orbits, which will
most likely lead to their ejection from the system after some time.  Interestingly, there is a reasonable number
of comets being scattered into orbits with inclinations of up to 60$^{\circ}$ (value measured at the end
of the integration) even for comets entering the system with an initial inclination of zero. 

\begin{figure}
\begin{center}
\includegraphics[width=8.4cm]{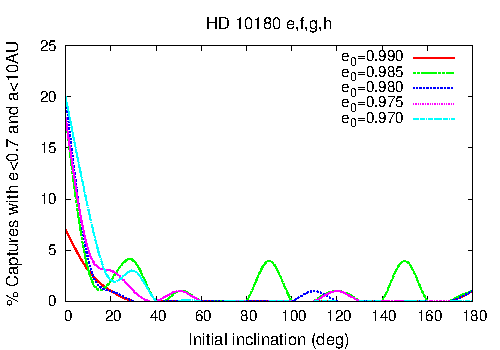}    % The printed column width is 8.4 cm.
\caption{HD~10180.  We show the probability (in \%) for a comet to be captured in an orbit with moderate values for
the semi-major axis and eccentricity.  The results depend on the comet's initial inclination and eccentricity.
We find that it is more probable for comets entering the system with low initial inclination to be captured
in a moderate orbit.} 
\label{fig:1.1}
\end{center}
\end{figure}

\begin{figure}
\begin{center}
\includegraphics[width=8.4cm]{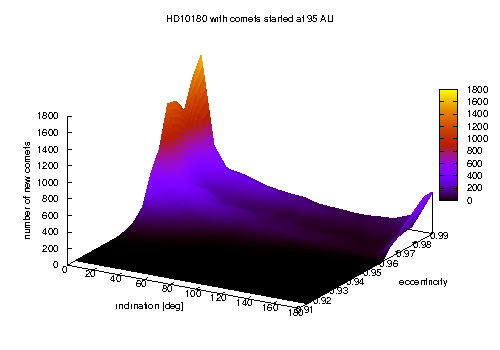}    % The printed column width is 8.4 cm.
\caption{HD~10180. The color coding shows the number of new comets inserted into the system.
This number also corresponds to the number of comets ejected from the system during the duration of the integration.
The higher initial eccentricity and the lower the initial inclination, the more likely it is for a comet to get ejected.}
\label{fig:1.2}
\end{center}
\end{figure}

\begin{figure}
\begin{center}
\includegraphics[width=8.4cm]{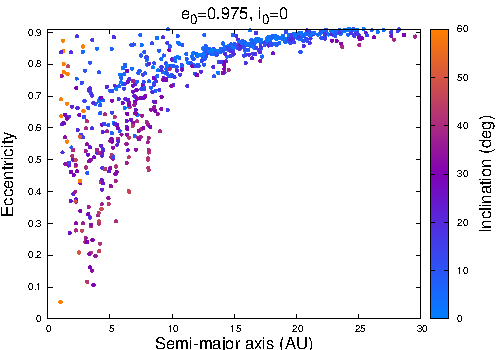}    % The printed column width is 8.4 cm.
\caption{HD~10180. Orbital properties of comets captured in the system after 1 Myr of integration.
The color coding indicates the inclination at the time of the capture. } 
\label{fig:1.3}
\end{center}
\end{figure}

\begin{figure}
\begin{center}
\includegraphics[width=8.4cm]{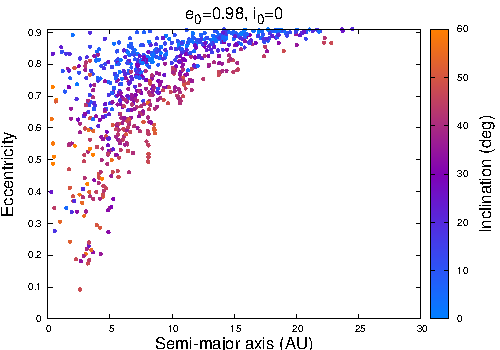}    % The printed column width is 8.4 cm.
\caption{HD~10180. Same as Figure~\ref{fig:1.3}, but for a slightly different initial eccentricity for the comets.} 
\label{fig:1.4}
\end{center}
\end{figure}

\subsection{47~UMa}

\subsubsection{The System}
\vspace*{0.2cm}
\noindent The system of 47~UMa shows significant similarities to the Solar System,
although the host star is more evolved along the main-sequence.  The stellar parameters are:
$T_\mathrm{eff}$=5890~K, $M$ = 1.03~$M_{\odot}$, and $R$ = 1.17~$R_{\odot}$, which have been
determined by \cite{Henry1997} and \cite{Gonzalez1998} with updates by \cite{Kovtyukh2003}
and \cite{vanBelle2009}.  Observations revealed three Jupiter-type planets with properties
shown in Table~\ref{t1}.  According to \cite{Kopparapu2013} the limits of the system's
general habitable zone (HZ) are given as 1.19~au and 2.04~au. The outer parts of the
radiative HZ is unavailable for hosting planets due to the influence of 47~UMa~b,
which is responsible for a truncation of the HZ at approximately 1.6~au.

\begin{table}
	\centering
\begin{tabular}{c|rlc}
\hline
Name & $a$ (au) & $e$ &  $m$ ($\mathrm{m_{Jupiter}}$) \\
\hline
47 UMa b & 2.1 & 0.032 &  2.53 \\
47 UMa c & 3.6 & 0.098 &  0.54 \\
47 UMa d & 11.6 & 0.16 &  1.64 \\
\hline
\end{tabular}
\caption{Properties of the planets in the system of 47~UMa. For details and updates see \protect\url{http://exoplanet.eu/}.}\label{t1}
\end{table}

For our simulations, besides the planets known in the 47~UMa system, we assumed additional Earth-mass planets
orbiting in the HZ in order to investigate the transport of water by comets to potentially habitable terrestrial planets.
These fictitious planets have been placed at 1.0~au, 1.25~au, and 1.584~au, respectively.  The latter planet is assumed to be
in 3:2 resonance with 47~UMa~b.

\subsubsection{Results}
\vspace*{0.2cm}
\noindent With respect to 47~UMa, we focused on the possibility of water transport by comets to the system's HZ.
Thus, we investigated the collisions of comets with fictitious Earth-mass planets with the known radius of the Earth. 
Figure~\ref{fig:2.1} and \ref{fig:2.2}
show collisions of comets with those bodies.  It is evident that the number of comets in each case decreases with
lower distance to 47~UMa~b.  The Hilda-type planet (originally placed at 1.584~au) did not undergo 
any collisions at all, which leads to the conclusion
that an object in resonance with 47~UMa~b is shielded from collisions by the massive planets. 

\begin{figure}
\begin{center}
\includegraphics[width=8.4cm]{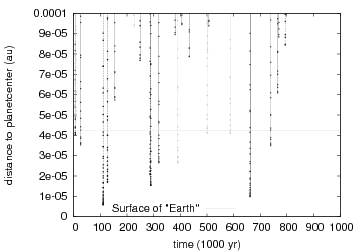}    % The printed column width is 8.4 cm.
\caption{47~UMa. The figures show cometary collisions with an Earth-mass planet placed at 1~au.
The green line marks the surface of the planet, assuming an Earth-like composition.} 
\label{fig:2.1}
\end{center}
\end{figure}

\begin{figure}
\begin{center}
\includegraphics[width=8.4cm]{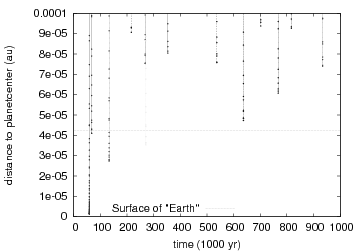}    % The printed column width is 8.4 cm.
\caption{47~UMa. Same as Figure~\ref{fig:2.1}, but for an Earth-mass planet placed at 1.25 au.} 
\label{fig:2.2}
\end{center}
\end{figure}

\begin{figure}
\begin{center}
\includegraphics[width=8.4cm]{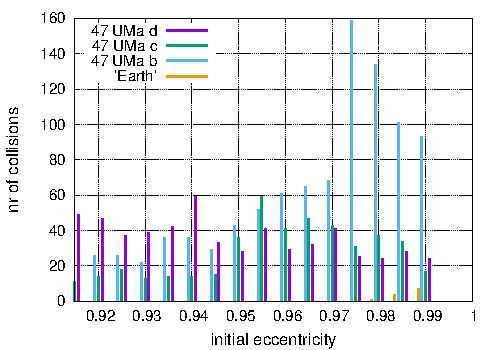}    % The printed column width is 8.4 cm.
\caption{47~UMa. The number of collisions depending on the initial eccentricity of the comets.
Only comets with initially high eccentricities are able to penetrate the inner planetary system.
Thus, 47~UMa~b and the Earth-mass planet (originally placed at 1~au) experience more (or only) collisions
with comets of initially higher eccentricity.} 
\label{fig:2.3}
\end{center}
\end{figure}

In Figure~\ref{fig:2.3} we show the number of collisions of comets with all planets in the 47~UMa system.
One can observe that the innermost planet 47~UMa~b receives the most collisions with comets with initially
high eccentricities.  Only those comets are able to come as close to the star to cross 47~UMa~b's orbit.
The fictitious Earth-mass planet positioned further-in undergoes only very few collisions. Thus,
the possibility of water transport by comets to terrestrial planets in 47~UMa's HZ is uncertain. 

\begin{figure}
\begin{center}
\includegraphics[width=8.4cm]{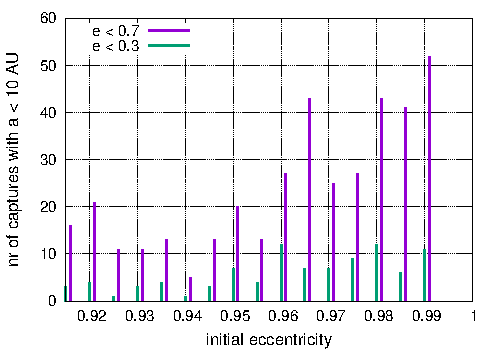}    % The printed column width is 8.4 cm.
\caption{47~UMa. The number of comets captured in orbits with moderate values for the semi-major axis and eccentricity.
The higher the initial eccentricity of the comets, the more probable it is for them to interact with the planets and
thus the more probable it is to be captured.} 
\label{fig:2.4}
\end{center}
\end{figure}

Figures~\ref{fig:2.4} and \ref{fig:2.5} depict the number of comets captured in orbits with moderate values for the
semi-major axis and eccentricity as a function of their initial eccentricity and inclination.
A repeating pattern emerges once again:  Comets with relatively high initial eccentricities are able to get closer
to the star and the inner planetary region.  This increases the likelihood of close encounters with one of the system planets,
allowing the comets to attain orbits with relatively small semi-major axes and eccentricities. Comets which plunge into
the system in the plane of the planetary orbits are also more likely to experience orbit-changing encounters
caused by one of the planets. Further results have been given by \cite{Cuntz2018}.

\begin{figure}
\begin{center}
\includegraphics[width=8.4cm]{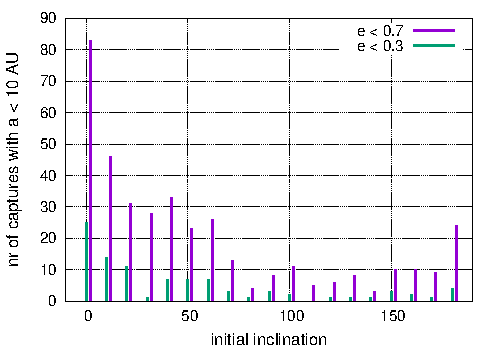}    % The printed column width is 8.4 cm.
\caption{47~UMa. Same as Figure~\ref{fig:2.4}, but for the initial inclination of the comets.
Comets entering the system in the plane have more time to interact with the planets; thus, it is more
probable for them to be captured in orbits with low values for the semi-major axis and eccentricity.} 
\label{fig:2.5}
\end{center}
\end{figure}

\subsection{HD~141399}

\subsubsection{The System}

\vspace*{0.2cm}
\noindent 
HD~141399 harbors 4 planets.  Its parameters are: $T_\mathrm{eff}$=5600~K and $M$ = 1.07~$M_{\odot}$;
thus, this star is somewhat cooler than the Sun but slightly more massive and has a metallicity of [Fe/H] = 0.35.
The properties of the known planets are given in Table~\ref{t2}.  The planets of the system have masses between
0.5 and 1.4 Jupiter masses.  The system's HZ extends from 1.04~au and 2.34~au \citep{Kopparapu2013};
this range corresponds to a domain between the planets HD~141399~c and HD~141399~d.

\begin{table}
	\centering
\begin{tabular}{c|llc}
\hline
Name & $a$ (au) & $e$ &  $m$ ($\mathrm{m_{Jupiter}}$) \\
\hline
HD 141399 b & 0.415 & 0.04 &  0.451 \\
HD 141399 c & 0.689 & 0.048 &  1.33 \\
HD 141399 d & 2.09 & 0.074 &  1.18 \\
HD 141399 e &   5.0  &  0.26   & 0.66  \\
\hline
\end{tabular}
\caption{Properties of the planets in the system of HD~141399. For details and updates see \protect\url{http://exoplanet.eu/}.}\label{t2}
\end{table}

\begin{figure}
\begin{center}
\includegraphics[width=8.4cm]{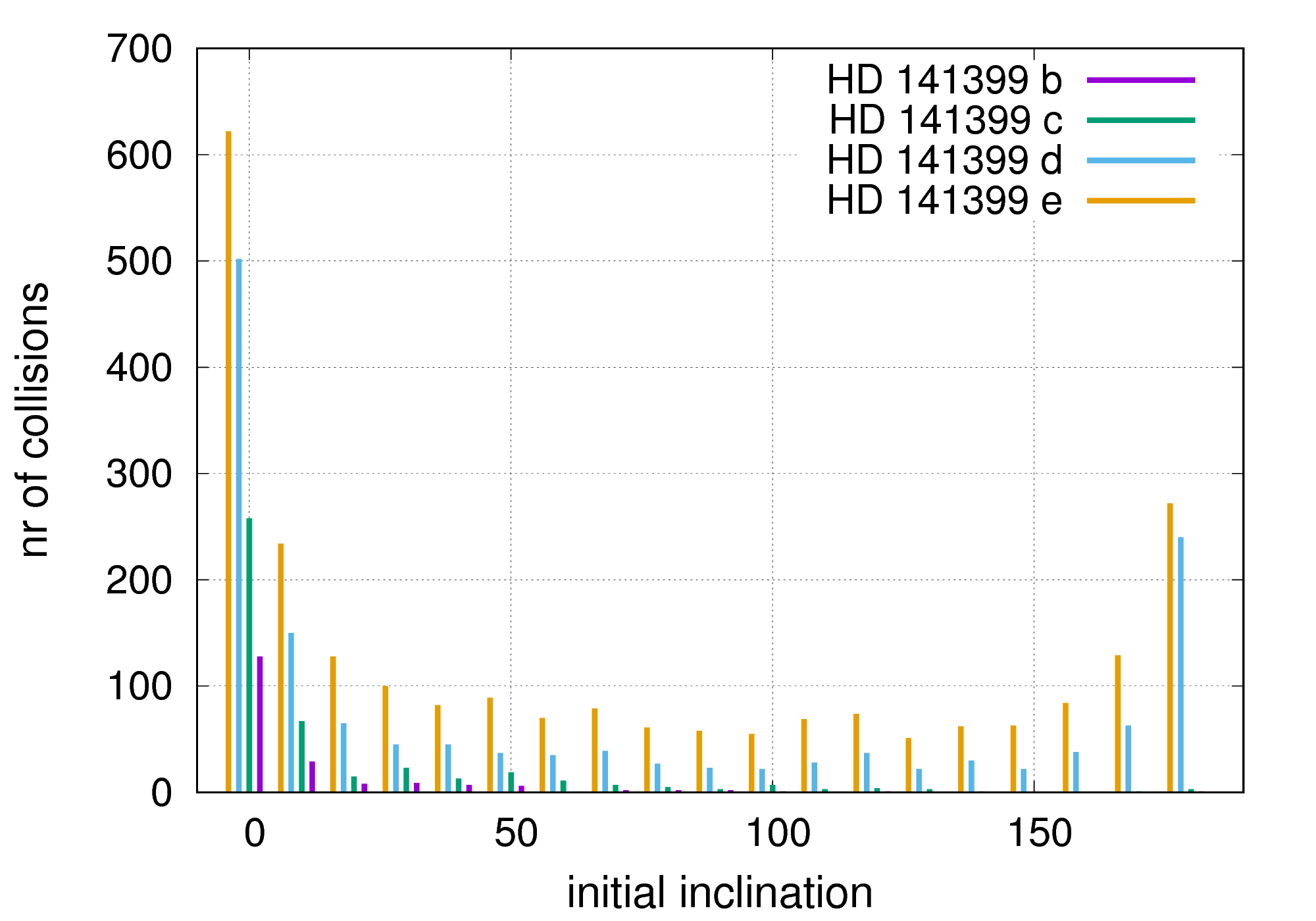}    % The printed column width is 8.4 cm.
\caption{HD~141399. We show the number of collisions with the different planets of the HD~141399 system
dependent on the initial inclination of the comets.  We depict the trajectories of all comets
with initial eccentricity between 0.92 and 0.99 and an initial semi-major axis of 100~au.} 
\label{fig:1}
\end{center}
\end{figure}

\subsubsection{Results}
\vspace*{0.2cm}
\noindent We investigated the interaction of comets assumed to originate from an Oort cloud analogue
with the four system planets of HD~141399.  The cometary cloud was set up as described in the previous
sections.  We monitored the paths of the comets on their way through the system and pursued a statistical
analysis of the captures, collisions, and escapes of the comets after their interaction with the planets.

Figure~\ref{fig:1} conveys the number of collisions with each planet.  One can observe that the outermost
planet HD~141399~e experiences most collisions with the cometary bodies.  This is accounted for with
the initial eccentricity of the comets entering the system.  Trajectories of comets with
initial eccentricities between 0.92 and 0.99 are included in our statistics.
Comets with relatively low eccentricities are unable to reach the inner planetary system without
undergoing close encounters and angular momentum exchange.  However, they are able to reach the orbit
of HD~131499~d.  Thus, the average number of comets crossing the orbit of the outermost planet is
greater than the number of comets reaching the orbit of one of the planets orbiting close to the star. 
Interestingly, the innermost planet HD~141399~b only encounters collisions with comets in prograde orbits
rather than comets with low inclinations ($i < 60^{\circ}$).  All in all, it is more likely for comets entering
the system in the ecliptic plane to collide with the planets than for comets entering the system
with relatively high inclinations. 

\begin{figure}
\begin{center}
\includegraphics[width=8.4cm]{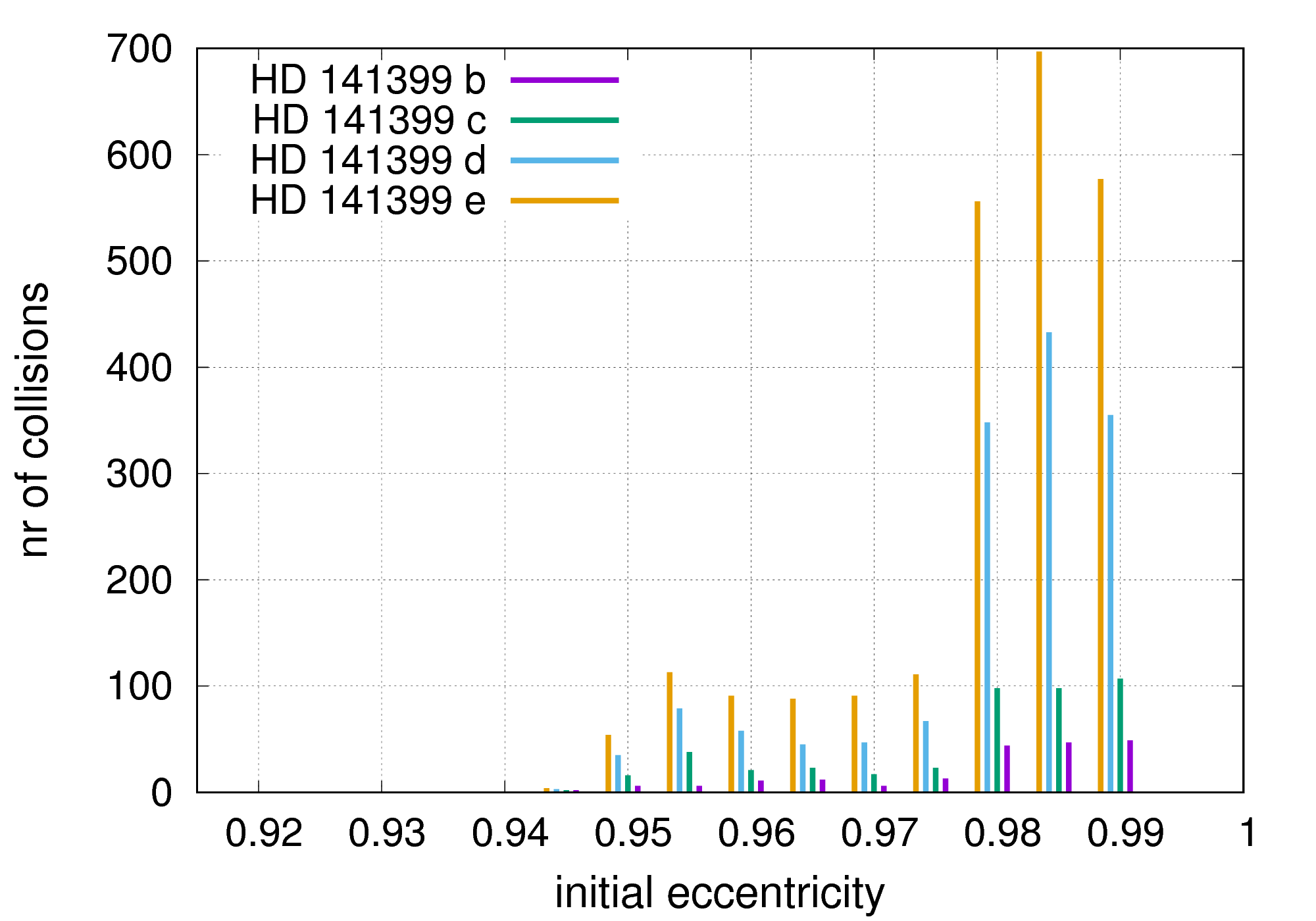}    % The printed column width is 8.4 cm.
\caption{HD~141399. Same as in Figure~\ref{fig:1}, but for the initial eccentricity of the comets.} 
\label{fig:2}
\end{center}
\end{figure}

Figure~\ref{fig:2} depicts the number of cometary collisions depending on the initial eccentricity for
comets with initial semi-major axes of 100~au.  Each bar includes the comets with all initial inclinations.
One finds essentially the same behavior as given in Figure~\ref{fig:1}: The outermost planet HD~141399~e
encounters most collisions with the comets entering the system.  Comets with initial eccentricities lower than
0.945 do not reach the orbit of even this planet; therefore, no collisions occur for those comets.  The higher
the eccentricity, the more collisions with the planets occur.  Interestingly, for comets with initial
eccentricities of $e > 0.98$, the number of collisions with the two outer planets drops while for the
two inner planets that value stays about the same.  This may be due to the trajectories of the comets.
Comets with high values for the eccentricity pass by the outer planets rather quickly as their perihelion
is farther in.  Thus, the time of interaction with the outer planets is relatively short, thus reducing
the probability of a collision with one of the planets.

\begin{figure}
\begin{center}
\includegraphics[width=8.4cm]{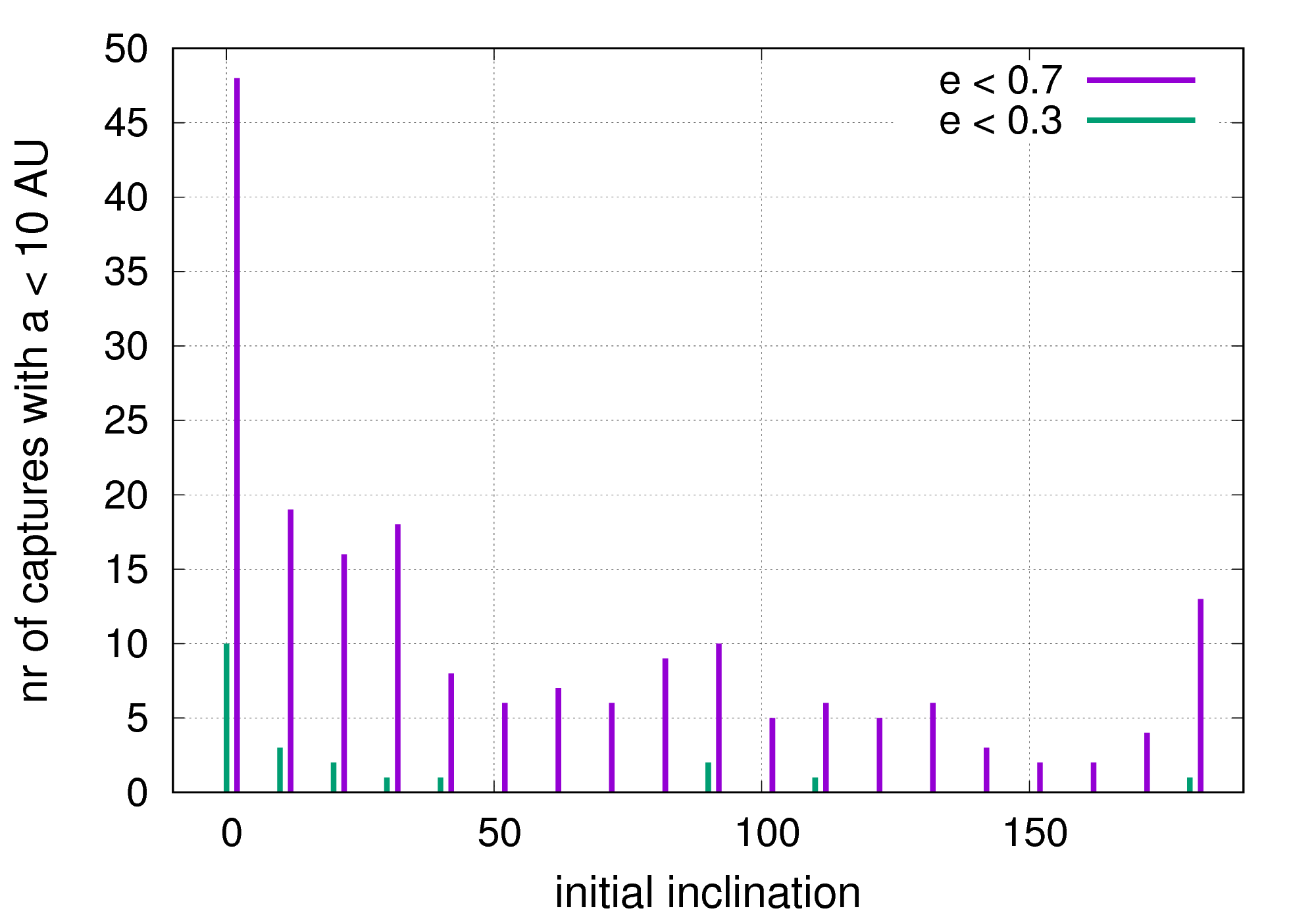}    % The printed column width is 8.4 cm.
\caption{HD~141399. Depiction of the number of captures of comets in orbits with low values for the semi-major axis and eccentricity.} 
\label{fig:3}
\end{center}
\end{figure}

Another outcome based on the exchange of angular momentum between the planets and comets entering the system of HD~141399
is the capture of a comet in a moderate orbit. We define moderate orbits as orbits with $a < 10$~au and
eccentricities lower than 0.7 (violet line in Figure~\ref{fig:3}). In special cases the eccentricities dropped to 0.3 (green lines). These orbits represent short-periodic comets. In this figure the number of comets fulfilling this criterion is given as a function of the initial inclination of the comets. Each bar includes comets with all initial eccentricities.  It is evident that the capture of a comet in such orbits
is most likely for comets entering the system in the ecliptic plane.  Interestingly, it is found that even comets with
initially high inclination can be captured in such orbits.  The likely underlying explanation considers the orbits
of the two inner giant planets.  The short-period orbits of HD~141399~b and HD~141399~c allow the planets
to effectively interact with the comets entering the system from above, which makes it more likely for them to be
scattered into orbits of small semi-major axes and low eccentricities.

\begin{figure}
\begin{center}
\includegraphics[width=8.4cm]{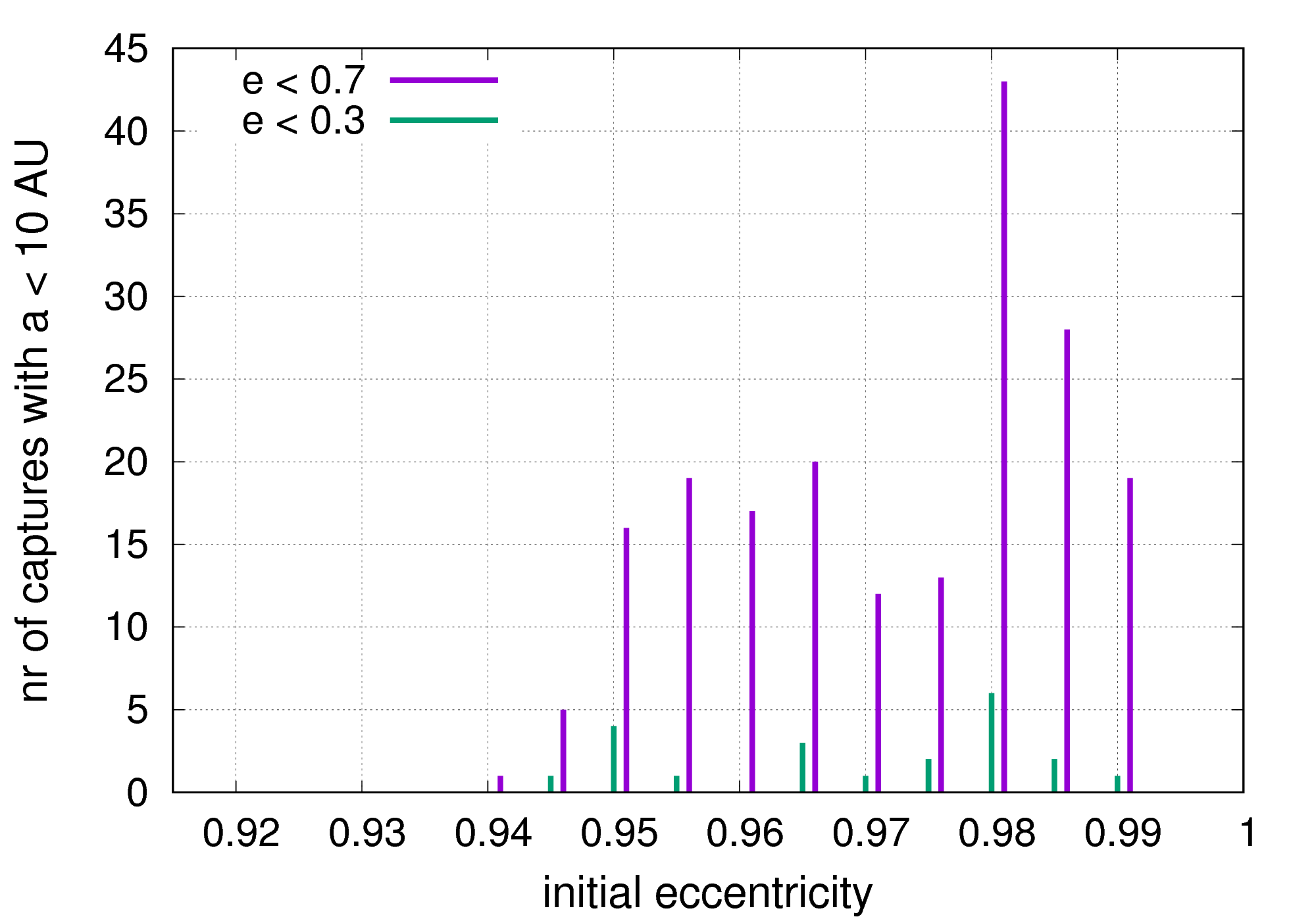}    % The printed column width is 8.4 cm.
\caption{HD~141399. Same as in Figure~\ref{fig:3}, but for the initial eccentricity of the comets.} 
\label{fig:4}
\end{center}
\end{figure}

Figure~\ref{fig:4} shows the same results but for the initial eccentricity of the comets.  As expected,
the number of comets captured into orbits with $a < 10$~au and low eccentricities is higher for comets
with initially high eccentricity.  The same reasons apply to the probability of collisions with the planets apply here.
Comets with initially high eccentricities are able to reach the planetary orbits and thus experience changes
in their angular momentum resulting in capture.  The same trend as given in Figure~\ref{fig:2} is here visible
as well: the probability for comets with initial eccentricity $e > 0.98$ drops. 

\begin{figure}
\begin{center}
\includegraphics[width=8.4cm]{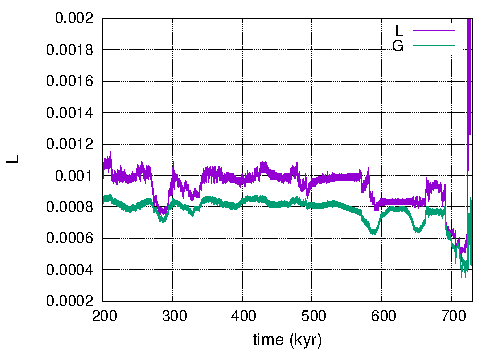}    % The printed column width is 8.4 cm.
\caption{HD~141399. Depiction of the Delaunay elements L and G of one comet experiencing a rather stable phase
in a captured orbit prior to experiencing a close encounter.} 
\label{fig:5}
\end{center}
\end{figure}

In Figure~\ref{fig:5} we depict the Delaunay elements L and G of one comet experiencing a type of stable phase
before undergoing a close planetary encounter and being scattered to a chaotic orbit; this behavior occurs at
about 730~kyr after the integration has been started.  L represents the evolution of the semi-major axis.
One can clearly identify the stable phases also visible in Figure~\ref{fig:6}.  Delaunay element G (i.e., 
the connection between $a$ and $e$) confirms that behavior.

The Delaunay elements are calculated as follows:

L = $\kappa$ $\cdot$ $\sqrt{a}$ \\
G = L $\cdot$ $\sqrt{1 - e^2}$ \\
H = G $\cdot$ $\cos(i)$ 

Figure~\ref{fig:6} shows the evolution of the semi-major axis of the same comet as depicted in Figure~\ref{fig:5}.
The figure shows stable phases of the comet, which is captured in a more or less stable orbit about the host star.
With $\kappa$ connected to the gravitational constant and the masses involved.

\begin{figure}
\begin{center}
\includegraphics[width=8.4cm]{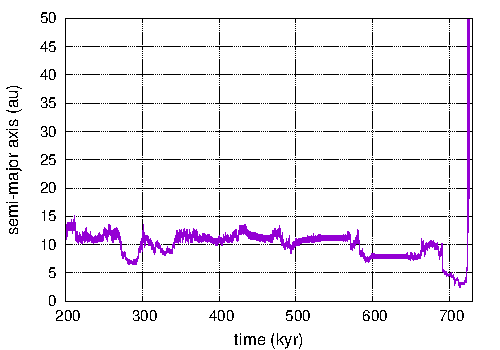}    % The printed column width is 8.4 cm.
\caption{HD~141399. Semi-major axis evolution of the comet depicted in Figure~\ref{fig:5}.  Phases of stability are clearly visible.} 
\label{fig:6}
\end{center}
\end{figure}

\section{Conclusion}

Since very recently we have evidence about comets in extrasolar planetary systems obtained
through spectroscopic observations.  The discovery of traces of objects, presumably created
through stellar interaction, was first reported some 20 years ago; however, theoretical work
predicting the existence of comets outside our Solar System has been presented even much earlier.
We scratched on the history of the discovery of exocomets since the end of the last century
and discussed the role of Oort clouds.  They are expected to exist around virtually every star
due to their formation history in large molecular clouds.  We then explored the possible dynamics
of comets penetrating the inner parts of extrasolar systems.  Thus, we discussed three specific
examples of extrasolar systems with gas giants in order to compare the outcomes with the cometary
structure in our Solar System.

Our goal was twofold.  First, we studied how the dynamical elements of comets arriving
from large distances are altered, perhaps resulting in the formation of families of comets akin
the Jupiter- and the Halley-cometary families in our Solar System.  Second, we planned to determine
the percentages of comets colliding with the planets or being ejected from the system because of
close encounters with one of the large planets.  However, in the three systems, which are:
HD~10180, 47~UMa, and HD~141399, we could not identify the formation of such families.  But
we found evidence for long-term captures of comets into orbits where they spent up to millions of
years in essentially stable orbits quite close to their host star.

For HD~10180, a tightly packed system with 6 or even 9 planets (including at least 5 Neptune-like giants),
we were able to show how the exocomets may have been captured or ejected from the system depending
on their initial eccentricities and inclinations.

For 47~UMa, a system containing three gas giants, we did the same type of investigation, but in addition
we checked how a possible Earth-mass planet could get its water inventory from comets.  Three different
cases were pursued: an Earth-mass planet placed at 1 au, another Earth-mass planet placed at 1.5 au,
and a third Earth-mass planet in the 3:2 MMR with the innermost system planet (47~UMa~b);
the habitable zone in 47~UMa is confined approximately between $\sim$0.9~au and $\sim$2.1~au for this late-stage main-sequence star.
The results show that there is little opportunity for the Earth-mass planets to collect water due to cometary collisions.
If water is present on any of those planets (if existing), an alternate mechanism of transport is required; see, e.g.,
\cite{Raymond2017} and references therein for proposed scenarios.

HD~141399, a star hosting four massive gas giants, was the final exoplanetary system of our investigations aimed at
studying cometary dynamics.  In all of our numerical experiments, the outermost planet, although the least massive,
suffered from many collisions, whereas the innermost giant planet had only very few ones.

\begin{figure}
\begin{center}
\includegraphics[width=0.3\textwidth, angle=270]{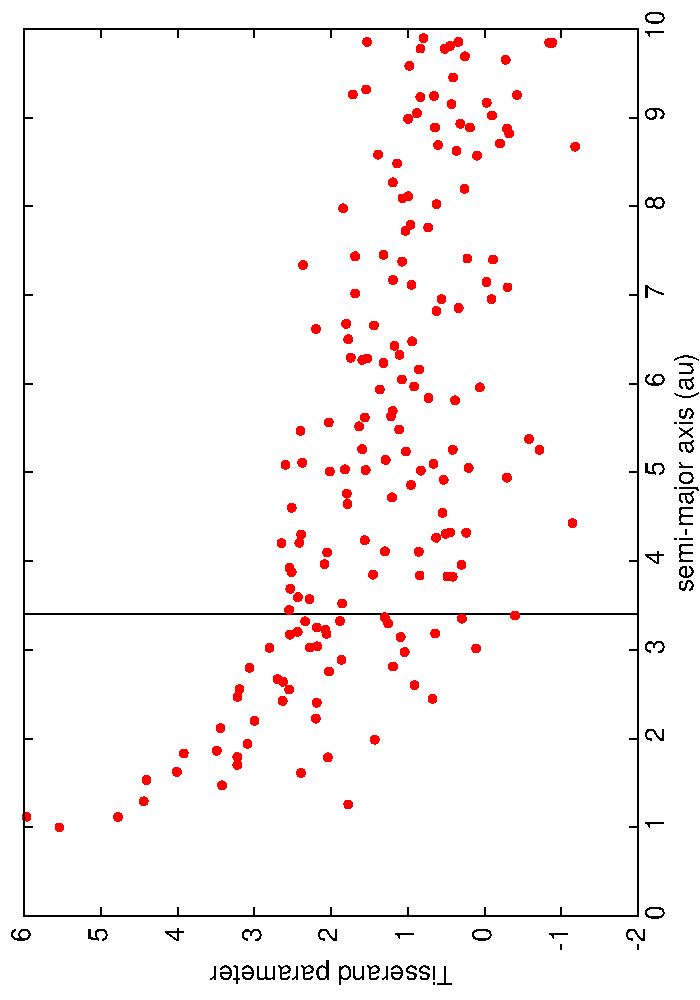}\\
\includegraphics[width=0.3\textwidth, angle=270]{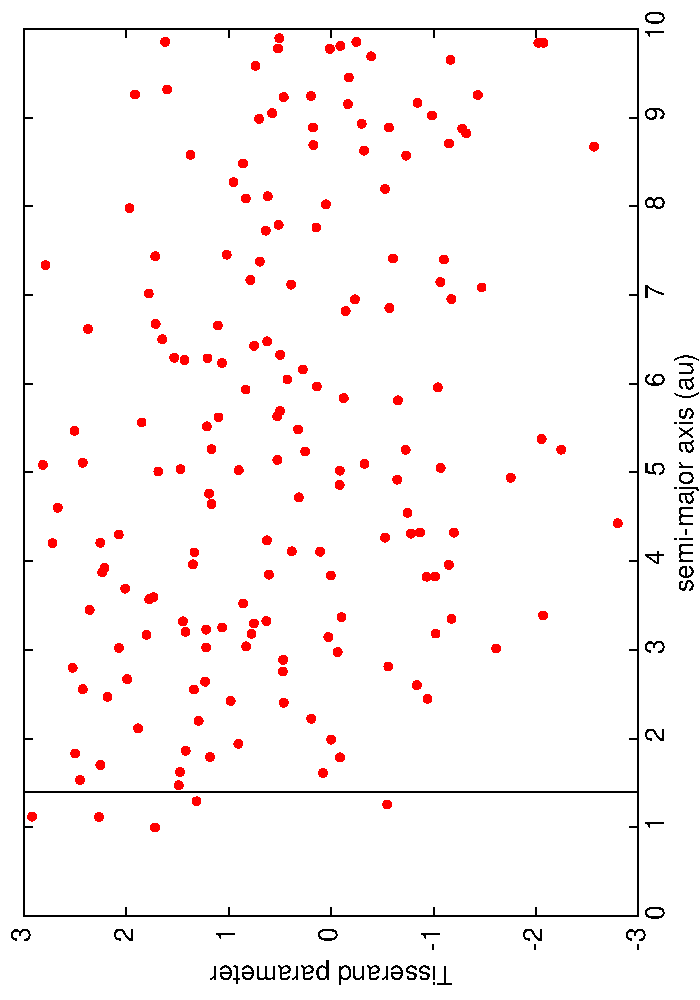}    % The printed column width is 8.4 cm.
\caption{Upper panel: Tisserand parameter versus semi-major axes for a gas giant at
$a=3.4$~au pertaining to HD~10108~h.  Lower panel: Same but for the planet HD~10108~g.
The vertical lines in both panels depict the location of the respective planets.} 
\label{fig:tiss}
\end{center}
\end{figure}

As a supplementary aspect of our study, we also checked the Tisserand parameter {\bf T} for the systems.
However, no suitable information could be obtained from these values depicted in figure~\ref{fig:tiss}.
It is well-known that in the Solar System Jupiter's role is dominant (\cite{Carusi1992}, \cite{Carusi1995}).
However, for HD~10108 the gas giant at $a=3.4$~au (planet h) cannot play a dominant role
due to the presence of another gas giant at $a=1.4$~au (planet g).  Hence, no cometary family akin
to the Jupiter family in the Solar System can therefore occur.  Furthermore,  in the other systems
investigated here the Tisserand parameter is not giving important information on the cometary orbits either.
Although the values interior to the location of planet h in the upper panel of Fig~\ref{fig:tiss} show
a similar signal like in the respective plot for Jupiter family comets (see Fig. 5 in \cite{Rickman2010}), 
the dispersion is too large to draw meaningful conclusions.  In the lower panel of Fig.~\ref{fig:tiss}
the Tisserand parameter (computed for planet g) is plotted versus the semi-major axis, but again 
no tendency can be seen due to the very large dispersion.

In the near future many more new observations are expected.  Some of them may be indicators of comets very
close to the host star. Thus, there will be a serious need for extending our numerical experiments to all systems with more
than one giant planet to acquire a more complete picture on the role of exocomets in proliferating water
to terrestrial planets.  For that we will carefully investigate possible terrestrial planets in the stellar habitable zones.
Our focus will be to determine the probability of collisions between comets, expected to originate from cometary clouds
at the outskirts of those systems, and terrestrial system planets, as indicated by theory or established by observations.

\begin{ack}
This research is supported by the Austrian Science Fund (FWF) through grant S11603-N16 (R.D. and B.L.).
Moreover, M.C. acknowledges support by the University of Texas at Arlington.  The computational results presented
have been achieved in part using the Vienna Scientific Cluster (VSC).
\end{ack}

\bibliographystyle{plainnat}
\bibliography{ifacconf}             % bib file to produce the bibliography

\appendix
%\section{A summary of Latin grammar}    % Each appendix must have a short title.
%\section{Some Latin vocabulary}              % Sections and subsections are supported  
                                                                         % in the appendices.
\end{document}